\documentclass[twocolumn]{aastex63}

\usepackage{bm} 
\usepackage[T1]{fontenc} 
\usepackage{subfigure} 
\usepackage{comment} 

\catcode`_ = \active
\newcommand_[1]{\sb{\mathrm{#1}}}

\usepackage{CJKutf8} 

\usepackage{ulem}

\hypersetup{linkcolor=red,citecolor=blue,filecolor=cyan,urlcolor=magenta}

\shorttitle{
matter density distributions of jets
}
\shortauthors{Ogihara, Ogawa \& Toma}

\graphicspath{{./}{figures/}}

\begin{document}
\begin{CJK*}{UTF8}{ipxm} 

\title{
Matter density distribution of general relativistic highly magnetized jets driven by black holes
}

\correspondingauthor{Taiki Ogihara}
\email{t.ogihara@astr.tohoku.ac.jp}

\author{Taiki Ogihara}
\affiliation{Astronomical Institute, Graduate School of Science, Tohoku University, Sendai, Miyagi, 980-8578, Japan}

\author{Takumi Ogawa}
\affiliation{Center for Computational Sciences, University of Tsukuba, Tsukuba, Ibaraki, 305-8577, Japan}
\affiliation{Astronomical Institute, Graduate School of Science, Tohoku University, Sendai, Miyagi, 980-8578, Japan}

\author{Kenji Toma}
\affiliation{Frontier Research Institute for Interdisciplinary Sciences, Tohoku University, Sendai, Miyagi, 980-8578, Japan}
\affiliation{Astronomical Institute, Graduate School of Science, Tohoku University, Sendai, Miyagi, 980-8578, Japan}


\begin{abstract}
High-resolution very long baseline interferometry (VLBI) radio observations have resolved the detailed emission structures of active galactic nucleus jets.
General relativistic magnetohydrodynamic (GRMHD) simulations have improved the understanding of jet production physics, although theoretical studies still have difficulties in constraining the origin and distribution of jetted matter.
We construct a new steady, axisymmetric GRMHD jet model to obtain approximate solutions of black hole (BH) magnetospheres, and examine the matter density distribution of jets. 
By assuming fixed poloidal magnetic field shapes that mimic force-free analytic solutions and GRMHD simulation results and assuming constant poloidal velocity at the separation surface, which divides the inflow and outflow,
we numerically solve the force-balance between the field lines at the separation surface and analytically solve the distributions of matter velocity and density along the field lines.
We find that the densities at the separation surface in our parabolic field models roughly follow $\propto r_{ss}^{-2}$ in the far zone from the BH, where $r_{ss}$ is the radius of the separation surface.
When the BH spin is larger or the velocity at the separation surface is smaller, the density at the separation surface becomes concentrated more near the jet edge.
Our semi-analytic model, combined with radiative transfer calculations, may help interpret the high-resolution VLBI observations and understand the origin of jetted matter.
\end{abstract}

\section{Introduction} \label{sec:intro}

Relativistic collimated outflows (or jets) are observed in active galactic nuclei (AGNs). 
They originate from the system composed of a supermassive black hole (BH), an accretion disk, and a surrounding gas at the center of the galaxy.
Very long baseline interferometry (VLBI) radio observations have revealed detailed emission structures of AGN jets. 
Limb-brightened structures have been observed at mm-cm wavelengths in jets of M87 \citep{Kovalev2007, Walker2008, Hada2011, Hada2016, Mertens2016, Kim2018, Walker2018}, Mrk 501 \citep{Piner2008}, Mrk 421 \citep{Piner2010}, Cyg A \citep{Boccardi2016}, and 3C84 \citep{Nagai2014, Giovannini2018}, and a triple-ridge structure composed of the limb-brightened components and an additional central bright component has been observed in the M87 jet \citep{Asada2016, Hada2017, Walker2018}.
The Event Horizon Telescope (EHT) has revealed the ring-like emission structure just around the BH horizon of M87 \citep{Event-Horizon-Telescope-Collaboration2019, Event-Horizon-Telescope-Collaboration2019a, Event-Horizon-Telescope-Collaboration2019b, Event-Horizon-Telescope-Collaboration2019c, Event-Horizon-Telescope-Collaboration2019d, Event-Horizon-Telescope-Collaboration2019e}.
These high-resolution observations have provided us hints for testing theoretical jet models.

The plausible driving mechanism of relativistic jets is the Blandford-Znajek (BZ) process \citep{Blandford1977}.
The frame-dragging effect on a rotating BH magnetosphere twists the poloidal magnetic field lines that thread the ergosphere and creates the outward Poynting flux extracting the energy and angular momentum from the BH \citep{Koide2002, Komissarov2004, Dermer2009, Toma2014, Toma2016, Kinoshita2018}.
General relativistic magnetohydrodynamic (GRMHD) simulations show that a highly magnetized region emerges around the rotational axis, where the BZ process works \citep[e.g.][]{McKinney2004, McKinney2009, Tchekhovskoy2011, Takahashi2016, Nakamura2018, Porth2019}. 
The Poynting flux can be converted to the matter kinetic energy flux, accelerating the matter to the relativistic speed \citep[e.g.][]{Komissarov2007, Komissarov2010, Lyubarsky2009, Tchekhovskoy2009, Beskin2010, Toma2013, Tanaka2020}.

A fundamental problem is the origin of the jetted matter. 
In the GRMHD simulations, the number density in the highly magnetized region becomes low around the `separation surface' between the inflow passing through the BH horizon and the outflow accelerating relativistically. 
The continuous injection of plasma is required to maintain a steady jet, although the large-scale magnetic field lines prevent thermal plasma from diffusing into the highly magnetized region.
Plasma could be replenished via the $e^+e^-$ pair-creation gap formation \citep[e.g.][]{Blandford1977, Beskin1992, Broderick2015, Hirotani2016, Levinson2017, Kisaka2020}, and/or the annihilation of high-energy photons from the accretion disk \citep{Levinson2011, Moscibrodzka2011, Kimura2020}.
However, these non-thermal effects are not considered in GRMHD simulations because of uncertainties in the physics of mass loading. 
Instead, they set the artificial density floor values to prevent the numerical difficulty \citep{McKinney2004, Riordan2018}. 

Solving the steady equations do not need the density floor. 
The mass density distribution of the solutions may constrain the mass-loading mechanisms.
The analytical solution of the steady axisymmetric GRMHD flow along a prescribed field line has been developed in the literature \citep{Bekenstein1978, Camenzind1986a, Takahashi1990, Globus2014}.
Once the field line configuration and the Bernoulli parameters are given, one can solve the Bernoulli equation (or wind equation) for each field line.
\citet{Pu2015} showed an analytic solution for the single field line which mimics the GRMHD simulation results \citep{Tchekhovskoy2010, Nakamura2018}. \citet{Pu2020} presented a wind solution, which successfully pass the fast points, for multiple field lines with the prescribed toroidal field, but they do not consider the force-balance between the field lines (or the Grad-Shafranov (GS) equation).
To solve the GS equation is computationally demanding \citep{Nitta1991, Fendt1997, Beskin2010, Nathanail2014, Pan2017, Mahlmann2018}.
Recently, \citet{Huang2019} and \citet{Huang2020} succeeded in constructing steady axisymmetric numerical solutions for the monopole and parabolic field line configurations by iteratively solving the wind equation and the GS equation.
In \citet{Huang2020}, they introduced the `loading zone'.
The inner boundary of the loading zone is the null-charge surface, and the outer boundary is the surface where the Bernoulli equation's solution of the outflow becomes $u_p=0$.
The outflow and inflow start at these surfaces.  
They assumed the fluid number flux per magnetic flux $\eta$ as constant for different field lines, and did not try to constrain the mass-loading mechanism.

One can also investigate the distribution of mass density as well as other physical quantities by comparing observations with the synthetic images derived by radiative transfer calculations in the jet models
\citep{Broderick2009, Porth2011, Dexter2012, Lu2014, Moscibrodzka2014, Moscibrodzka2016, Moscibrodzka2017, Dexter2016, Jimenez-Rosales2018, Kawashima2019, Chael2018, Chael2019, Davelaar2019, Chatterjee2020, Jeter2020}.
Based on special relativistic force-free steady jet models, \citet{Takahashi2018} showed that a rapidly rotating BH accelerates the flow efficiently with remaining small toroidal velocity, leading to symmetric limb-brightened structure, which is compatible with the large-scale observations of M87 jet, and \citet{Ogihara2019} showed that the velocity field structure naturally produces the observed central ridge emission of M87 jet by relativistic beaming effect \citep[see also][]{Chernoglazov2019}.
They also demonstrated that the synthetic images of large-scale jets strongly depend on the density distribution of jets near the BHs. 
\citet{Kawashima2020} performed GR radiative transfer calculations to show that the emission at the bottom of the separation surface can reproduce the ring-like image just around the M87 BH.
Future EHT observations of M87 will unveil the emission structure between the limb-brightened one and the ring-like image \citep{Hada2019}. Parametric studies of steady GRMHD jet models and their comparison with the observations will extract the mass density distribution and other specific conditions around the separation surface.

In this paper, 
we introduce a new way to construct a steady axisymmetric GRMHD approximate solution to examine the jet's density distribution without being suffered from the density floor problem.
We fix the field line configuration which mimics the ones of force-free or highly-magnetized GRMHD simulation results \citep{Tchekhovskoy2008, Tchekhovskoy2010, Nakamura2018, Porth2019} with an additional term for obtaining trans-fast-magnetosonic solutions \citep[c.f.][]{Beskin2006, Pu2015}.
We numerically solve the transverse force-balance between the field lines at the separation surface to determine the distribution of the Bernoulli parameters, which include $\eta$, and analytically solve the Bernoulli equation along each field line.
We assume the poloidal velocities at the separation surface as constant for different field lines, and our model does not employ the `loading zone.'
This paper is organized as follows.
In Section \ref{sec:model}, we outline governing equations of steady, axisymmetric, cold, ideal GRMHD flows and our method to obtain approximate solutions.
We present calculation results in Section \ref{sec:results}.
In Section \ref{sec:split-monopole configuration case}, we first confirm that our method can approximately reproduce the split-monopole force-free solution of slowly rotating Kerr BH magnetosphere, and then in Section \ref{sec:parabolic configuration case}, we perform calculations for the parabolic field model.
We discuss the radial profiles of the density at the separation surface and compare our results with a mass loading mechanism and other studies in Section \ref{sec:discussion}. 
Summary and prospects are presented in Section \ref{sec:summary}.

\section{Model} \label{sec:model}

To examine the density distribution inside the jet, we study the general relativistic equations for steady, axisymmetric, cold ideal MHD flows.
We analytically solve the dynamics parallel to the prescribed poloidal field lines and numerically keep the transverse force balance at the separation surface. 

The spacetime geometry is given by the Kerr metric in the Boyer-Lindquist coordinates,
\begin{eqnarray}
    ds^2 = &-&\left(1-\frac{2r}{\Sigma}\right) dt^2 - \frac{4ar\sin^2\theta}{\Sigma} dt d\phi + \frac{\Sigma}{\Delta} dr^2 \nonumber \\ 
    &+& \Sigma d\theta^2 + \frac{[(r^2+a^2)^2-\Delta a^2\sin^2\theta]\sin^2\theta}{\Sigma} d\phi^2, \nonumber \\
\end{eqnarray}
where $a$ is the dimensionless spin parameter, $\Sigma=r^2+a^2\cos^2\theta$, and $\Delta=r^2-2r+a^2$.
Hereafter we use the unit of $c = G = 1$ and the BH mass $M = 1$. The radius of the event horizon is $r_{\rm H} = 1 + \sqrt{1-a^2}$, and the angular velocity of the BH is $\omega_{\rm H} = a/2r_{\rm H}$.

\subsection{Bernoulli equation}
The basic equations are the energy-momentum equation, the Maxwell equations, and the mass flux conservation law with the steady, axisymmetric, ideal MHD conditions.
They lead to the four integral constants along a field line (i.e. the Bernoulli constants), which are the total energy flux per particle $E$, the total angular momentum flux per particle $L$, the number density flux per magnetic flux $\eta$, and the so-called the `angular velocity of the magnetic field' $\Omega_{\rm F}$ \citep{Bekenstein1978}. They are given as follows:
\begin{eqnarray}
    &\hat E(\Psi)& = -u_0 - \frac{\Omega_F B_3}{4\pi\mu\eta}, \label{eq:EHAT} \\
    &\hat L(\Psi)& =  u_3 - \frac{B_3}{4\pi\mu\eta}, \label{eq:LHAT} \\
    &\eta(\Psi)& = -\frac{n u_1}{B_1} G_t = -\frac{n u_2}{B_2} G_t, \label{eq:eta}\\
    &\Omega_F(\Psi)& = \frac{F_{01}}{F_{13}} = \frac{F_{02}}{F_{23}}. \label{eq:OMEGAF}
\end{eqnarray}
Here we have introduced the magnetic flux function $\Psi(r,\theta)$ and the electromagnetic tensor $F_{\mu\nu}$. 
For the axisymmetric field, $F_{A3} = \partial_A \Psi (A=1,2)$. 
Then the magnetic field is given by
$B_\mu = 1/2 \epsilon_{\nu\mu\lambda\sigma} \xi^\nu F^{\lambda\sigma}$, where 
$\xi^{\mu}=(1,0,0,0)$ is the time-like Killing vector, $\epsilon_{\mu\nu\lambda\sigma}=\sqrt{-g}[\mu\nu\lambda\sigma]$,
$[\mu\nu\lambda\sigma]$ is the permutation symbol, and $g \equiv \det (g_{\mu\nu})$. 
$F_{A0} = E_A$ is the electric field.
$n$ is the fluid-frame number density of the matter, $u^\mu$ is the four-velocity, and $\mu$ is the specific enthalpy.
In the cold limit, $\mu$ is the rest energy of the plasma particle. 
$\hat E = E/\mu$, $\hat L = L/\mu$, and $G_t=g_{00} + \Omega_F g_{03}$. 

Combination of Equations (\ref{eq:EHAT}), (\ref{eq:LHAT}), (\ref{eq:eta}), and (\ref{eq:OMEGAF}) is reduced to the Bernoulli equation, i.e., a fourth-order equation of the poloidal velocity $u_p = \sqrt{u^1u_1 + u^2u_2}$,
\begin{equation}
    \sum_{{\it i}=0}^4 A_{\it i} u_p^i = 0, \label{eq:fourth-order up}
\end{equation}
where 
\begin{eqnarray}
    A_4 &=& 1, \nonumber \\ 
    A_3 &=& \frac{k_0 B_p}{2\pi \mu\eta G_t}, \nonumber \\ 
    A_2 &=& 1+\hat E^2 k_4 + \left( \frac{k_0 B_p}{4\pi \mu\eta G_t} \right)^2, \nonumber \\ 
    A_1 &=& \frac{B_p(k_0-\hat E^2 k_2)}{2\pi \mu\eta G_t}, \nonumber \\ 
    A_0 &=& k_0 (k_0-\hat E^2 k_2) \left( \frac{B_p}{4\pi \mu\eta G_t} \right)^2.
\end{eqnarray}
Here, $k_0 = - (g_{00} + 2\Omega_F g_{03} + \Omega_F g_{33})$, $k_2 = (1 - \hat L \Omega_F / \hat E)^2$, $k_4 = [ g_{33} + 2 g_{03} \hat L / \hat E + g_{00} (\hat L / \hat E)^2 ]/( g_{03}^2 - g_{00} g_{33} )$, and $B_p = \sqrt{B_1 B^1 + B_2 B^2}$ is the poloidal magnetic field. 
Equation (\ref{eq:fourth-order up}) is identical with Equation (34) in \citet{Pu2015} and Equation (23) in \citet{Huang2019}. 

Given the integral constants $\{\hat{E}(\Psi), \hat{L}(\Psi), \eta(\Psi), \Omega_F(\Psi)\}$ and the field line configuration $\Psi(r,\theta)$, one can solve the Bernoulli equation.
Then one can derive the number density distribution along the field line from the definition of $\eta$ (Equation \ref{eq:eta}).
The toroidal field can be calculated by
\begin{equation}
    B_3 = -4 \pi \mu\eta \frac{G_\phi \hat E + G_t \hat L}{M^2 - k_0},
    \label{eq:BB3}
\end{equation}
with the Alfven Mach number $M^2 = \pi \mu \eta^2 /n$ and $G_\phi = g_{03} + \Omega_F g_{33}$.

\subsection{Flux function model}
We assume that
$\Psi(r,\theta)$ has a form of 
\begin{equation}
    \Psi(r,\theta) = C \left[ \left( \frac{r}{r_H} \right)^\nu(1-\cos\theta) + \frac{1}{4} \epsilon r \sin\theta \right], 
\end{equation}
where 
$C$ is the normalization factor setting $\Psi(r_H,\pi/2)=1$.
$\nu$ and $\epsilon$ are the model parameters controlling the poloidal field line shape.
For $\epsilon=0$, the field line configurations of $\nu=0$ and $\nu=1$ have the monopole and parabolic shape in the far zone, respectively. 
$\Psi(r,\theta)$ with $\nu=0$ and $\epsilon=0$ is the exact solution of the force-free magnetosphere in a Schwartzschild spacetime \citep{Blandford1977}, and $\Psi(r,\theta)$ with $\nu=1$ and $\epsilon=0$ is the dominant term of the exact solution of the force-free magnetosphere in a Schwarzschild spacetime \citep{Blandford1977, Lee2004}.
The jet-disk boundary in GRMHD simulations defined as the magnetic-to-matter energy flux ratio $\sigma = |-\Omega_F B_3/(4\pi\mu\eta u_0)| = 1$ matches the parabolic configurations with a constant $\nu$ in a large computational domain \citep{Nakamura2018, Porth2019}.
$\epsilon$ represents a small disturbance from the force-free field lines, which makes the outflow accelerate by converting the Poynting flux to the kinetic energy flux \citep[c.f.][]{Beskin2006, Pu2015}.
When $\epsilon=0$, $B_p r^2\sin^2\theta$ becomes constant along the field line in a far zone \citep[][]{Pu2020}.
For the outflow to pass through the fast magnetosonic point, $B_p r^2\sin^2\theta$ needs to decrease with the radius around the fast magnetosonic point \citep{Begelman1994, Beskin2010, Toma2013}.
When $\epsilon > 0$, the field lines are collimated and $B_p r^2\sin^2\theta$ decreases. 
Then, the outflow can pass through the fast magnetosonic point.

\begin{table}
\centering
    \begin{tabular}{l|RRRRRR}
        model & \nu & \epsilon & a & \Omega_F(\Psi=1) & u_{p,ss} & \hat E_0 \\
        \hline 
        M0 & 0 & 0 & 0.1 & 0.4999\Omega_H & 10^{-3} & 10^3 \\
        \hline
        P1 & 1 & 10^{-4} & 0.9 & 0.35\Omega_H & 10^{-3} & - \\
        P2 & 1 & 10^{-4} & 0.8 & 0.35\Omega_H & 10^{-3} & - \\
        P3 & 1 & 10^{-4} & 0.95 & 0.35\Omega_H & 10^{-3} & - \\
        P4 & 1 & 10^{-4} & 0.9 & 0.35\Omega_H & 6 \times 10^{-4} & - \\
        P5 & 1 & 10^{-4} & 0.9 & 0.35\Omega_H & 1.4 \times 10^{-3} & - 
    \end{tabular}
    \caption{Parameter values used in our models.
    }
    \label{tab:parameter list}
\end{table}

\begin{figure}
    \centering
    \plotone{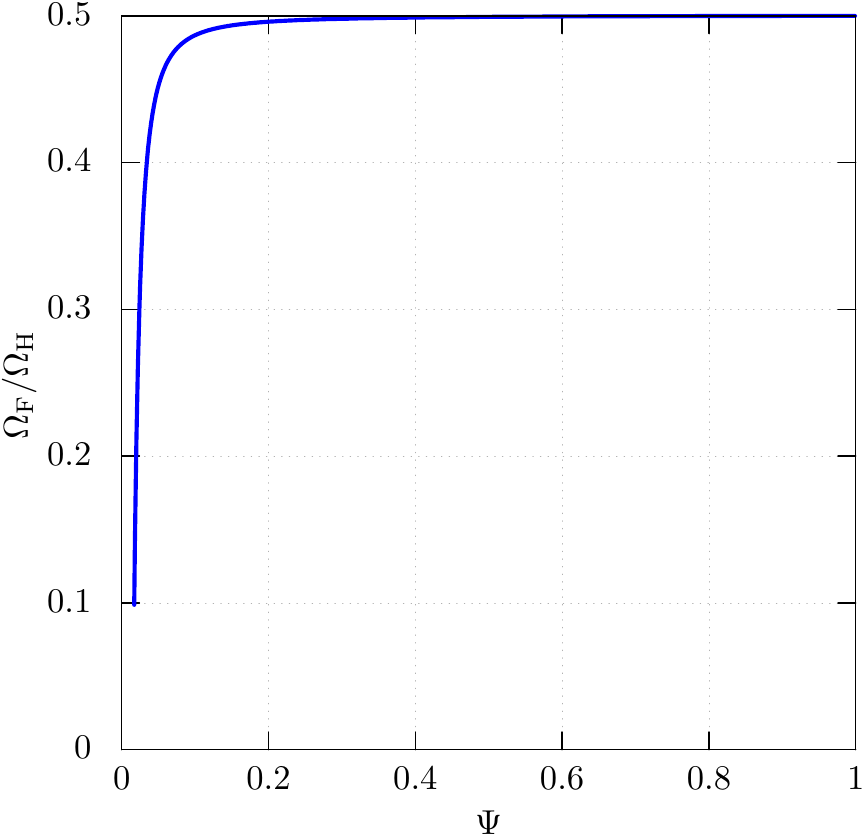}
    \plotone{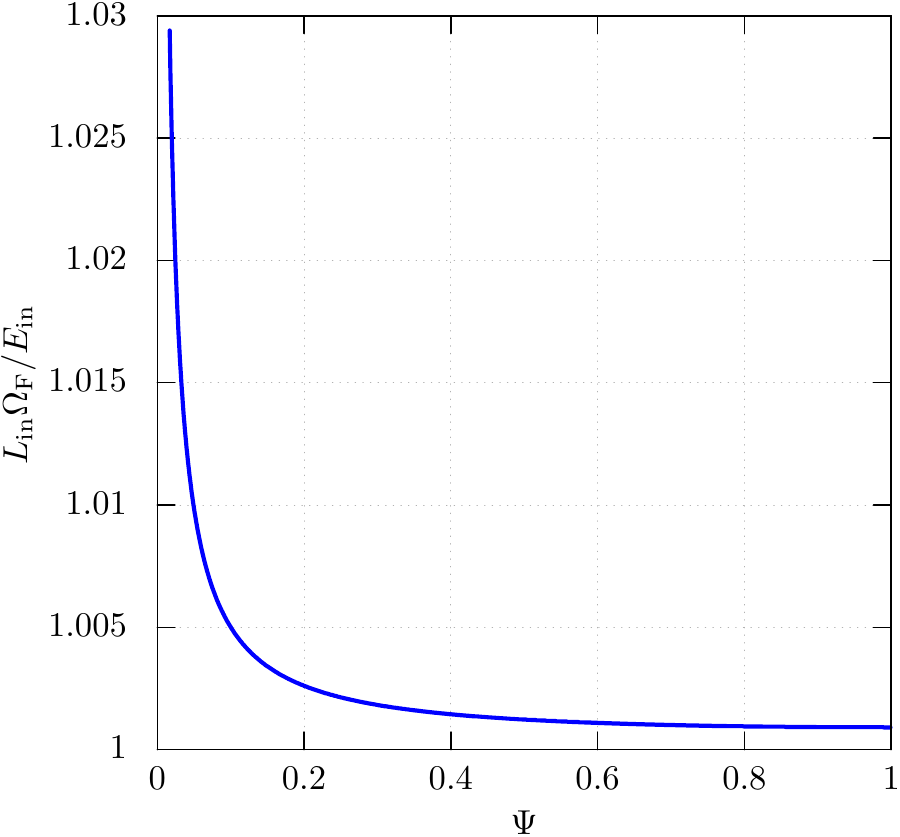}
    \caption{$\Omega_F(\Psi)$ and $L_{in} \Omega_F/E_{in} (\Psi)$ of the M0 model. Our result satisfies the conditions of the force-free split-monopole magnetosphere, $\Omega_F=0.5\Omega_H$ and $L \Omega_F/E=1$, within 1\% accuracy in $\Psi>0.1$. }
    \label{fig:monopole ss} 
\end{figure}

\begin{figure}
    \centering
    \plotone{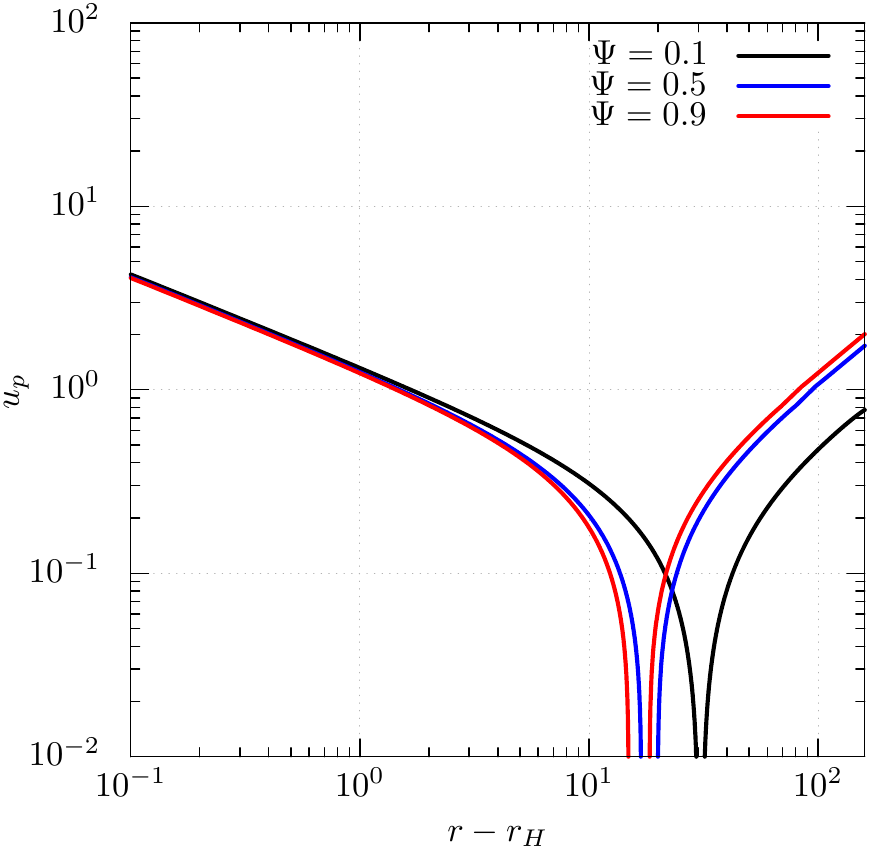}
    \plotone{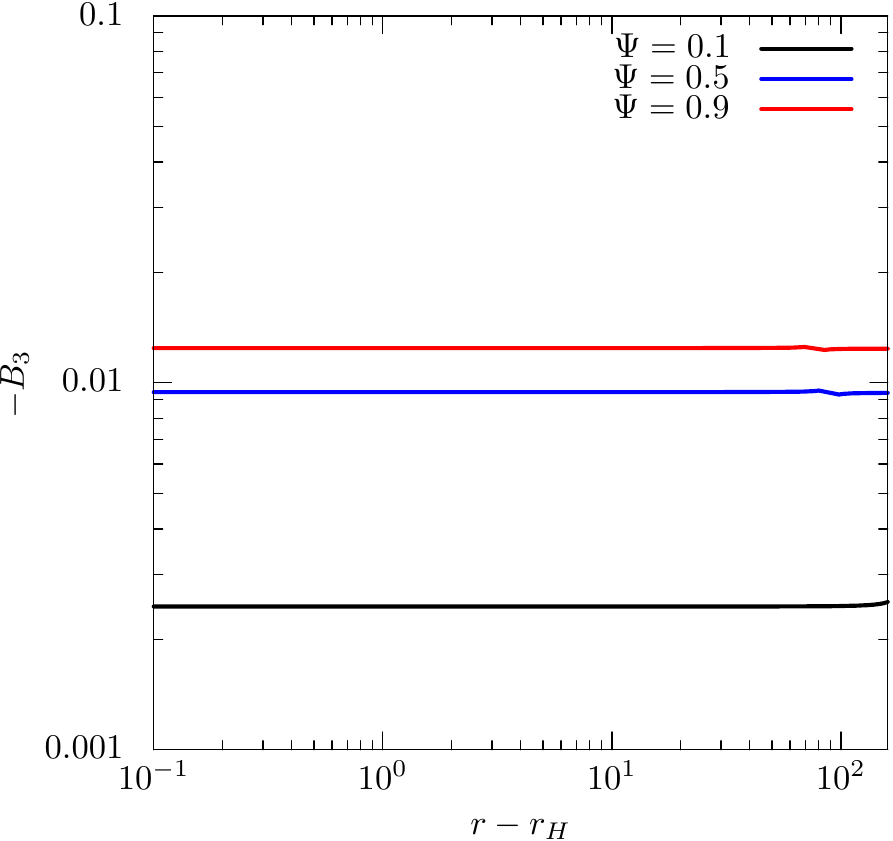}
    \caption{$u_p$, $B_3$ along the field lines $\Psi=0.1, 0.5,$ and $0.9$ of the split-monopole configuration model.
    We note that $u_p$ is the absolute value of the poloidal velocity.
    The constant $B_3$ along the field line represents that the energy conversion from the Poynting flux to the kinetic energy flux is inefficient in this model.}
    \label{fig:monopole wind}
\end{figure}

In this paper, we consider two models, a split-monopole configuration model ($\nu=0, \epsilon=0$), and a perturbed parabolic configuration model ($\nu=1, \epsilon= 10^{-4}$).
The aim of performing the split-monopole model is to check if our method can approximately reproduce the split-monopole force-free solution of slowly rotating Kerr BH magnetosphere, in which $\Omega_{\rm F} \approx 0.5\Omega_{\rm H}$ and $\hat{E}(\Psi) \propto \sin^2\theta$  \citep{Blandford1977}.
We set $a=0.1$ because such a small BH spin does not change the field line structure significantly from the monopole force-free solution near the horizon \citep{Tanabe2008, Tchekhovskoy2010}.
After confirming that our model can reproduce the split-monopole analytic solution well,
we perform a calculation of a perturbed parabolic magnetosphere around a rapidly rotating BH ($a=0.9$) to investigate the density distribution in the jet.

We list the parameter values used in our models in Table \ref{tab:parameter list}.
The M0 model is the split-monopole configuration model, and the results are shown in Section \ref{sec:split-monopole configuration case}.
The P1 model is our fiducial model in the parabolic configuration models.
We analyze the details of the P1 model in Section \ref{sec:parabolic configuration case}. 
The other parabolic configuration models (P2, P3, P4, and P5) are also analyzed to investigate the parameter dependence.

\subsection{Determining the Bernoulli parameters} \label{sec:Determining the Bernoulli parameters}

Solving the wind equation (Equation \ref{eq:fourth-order up}) requires the Bernoulli parameters $\{\hat E(\Psi), \hat L(\Psi), \Omega_F(\Psi), \eta(\Psi) \}$.
We determine $\hat L(\Psi), \eta(\Psi),$ and $\Omega_F(\Psi)$ in the same method for both of the field line configuration models, and determine $\hat E(\Psi)$ differently.

$\hat L(\Psi)$ and $\eta(\Psi)$ are determined simultaneously from the assumed poloidal velocity at the separation surface and the Znajek condition \citep{Znajek1977},
\begin{eqnarray}
    u_p(r_{ss}, \theta) &=& u_{p,ss} = \rm const. \\    
    B_3(r_H, \theta) &=& - \left[ \left( \frac{g_{33}}{g_{22}}\right)^{1/2} (\omega_H - \Omega_F) \partial_\theta \Psi \right]_{r=r_H},
    \label{eq:B3}
\end{eqnarray}
where $r_{ss}$ denotes the radius of the separation surface.
Combining $u_{p,ss}$ and the Znajek condition with Equation (\ref{eq:fourth-order up}) at the separation surface and Equation (\ref{eq:BB3}) at the horizon, one can obtain the two equations for solving $\hat L$ and $\eta$.
We set $u_{p,ss}= 10^{-3}$ for the M0, P1, P2, and P3 model.
The P2 and P3 models are used to investigate the BH spin dependence, while
the P4 and P5 models are used to investigate the $u_{p,ss}$ dependence.
The separation surface is defined by $\partial_p k_0 = 0$, where $\partial_p \equiv B^1 \partial_1 + B^2 \partial_2 $ is the differentiation in the direction along the field line.

$\Omega_F(\Psi)$ is determined to satisfy the force balance between field lines at the separation surface.
The component of the equation of motion perpendicular to the field line is 
\begin{equation}
    e_{(n)A}[\rho u^\nu u^A_{;\nu} - F^{A\nu}I_\nu] = 0 \label{eq:trans-field EoM},
\end{equation}
where $e_{(n)A}$ is the unit vector in the direction perpendicular to the field line in the poloidal plane $e_{(n)A}=E_A/\sqrt{E^B E_B}$ ($A,B=1,2$). 
We assume $\Omega_F(\Psi=1)=0.4999\Omega_H$ for the M0 model, and $\Omega_F(\Psi=1)=0.35\Omega_H$ for the parabolic configuration models.
Then we choose $\Omega_F(\Psi)$ by minimizing the left-hand side of Equation (\ref{eq:trans-field EoM}).
We do this only at the separation surface to obtain the approximate solutions. We confirm that this method works well for the M0 model and then apply it to the parabolic configuration models. We also evaluate how well our approximate solutions keep the transverse balance at the regions other than the separation surface.

Regarding of $\hat E(\Psi)$, we take different treatments for the split-monopole and parabolic cases.
\begin{description}
\item[(i)] For the M0 model, we set $\hat E = \hat E_0 \sin^2\theta$ and $\hat E_0 = 10^3$, following the dependence of the Poynting flux in the force-free monopole magnetosphere \citep{Blandford1977}. 
The small value of $u_{p,ss}$ is taken for the flow to be Poynting-flux dominated.
\item[(ii)] For the parabolic configuration models,
we determine $\hat E$ for the outflow to accelerate smoothly passing through the fast magnetosonic point. 
If $\hat E$ is too low, $u_p$ diverges after passing the Alfven point, 
while if $\hat E$ is too large, $u_p$ does not diverge, but $M^2$ does not reach $M^2_{fast} \equiv k_0 + (G_t^2 B_3^2)/(\rho_w^2 B_p^2)$, where $M^2 = M^2_{fast}$ should be satisfied at the fast magnetosonic point \citep[][]{Takahashi1990}.
We adjust $\hat E$ so that $u_p$ does not diverge and there is a point where $|1-M^2/M^2_{fast}| < 10^{-3}$.

We iteratively adjust $\hat E$ and $\Omega_F$ to obtain their values satisfying both of the condition for smooth acceleration and the force balance at an expected precision. 
\end{description}

From Equation (\ref{eq:eta}), $\eta$ has the opposite sign for the inflow and outflow. 
The sign of the second term of Equation (\ref{eq:EHAT}) and (\ref{eq:LHAT}) change between the inflow and outflow. 
On the other hand, the sign of the first term $u_0$ does not change, and $u_0$ is almost $-1$ at the separation surface.
Due to this, $\hat E$ and $\hat L$ must have different values for the inflow and outflow to keep the electromagnetic field smooth and continuous at the separation surface.
Thus, we set $\hat E_{\rm in} = - \hat E_{out} +2$, where $\hat E_{\rm in}$ and $\hat E_{\rm out}$ are the values for inflow and outflow, respectively.

\section{results} \label{sec:results}
In Section \ref{sec:split-monopole configuration case}, we perform the split-monopole configuration model and confirm that our model, which takes the trans-field balance only at the separation surface, can make the electromagnetic field consistent with the force-free solution.
In Section \ref{sec:parabolic configuration case}, we perform the parabolic configuration model and show the density distribution inside the jet.

\begin{figure*}
    \centering
    \includegraphics[height=0.4\textwidth]{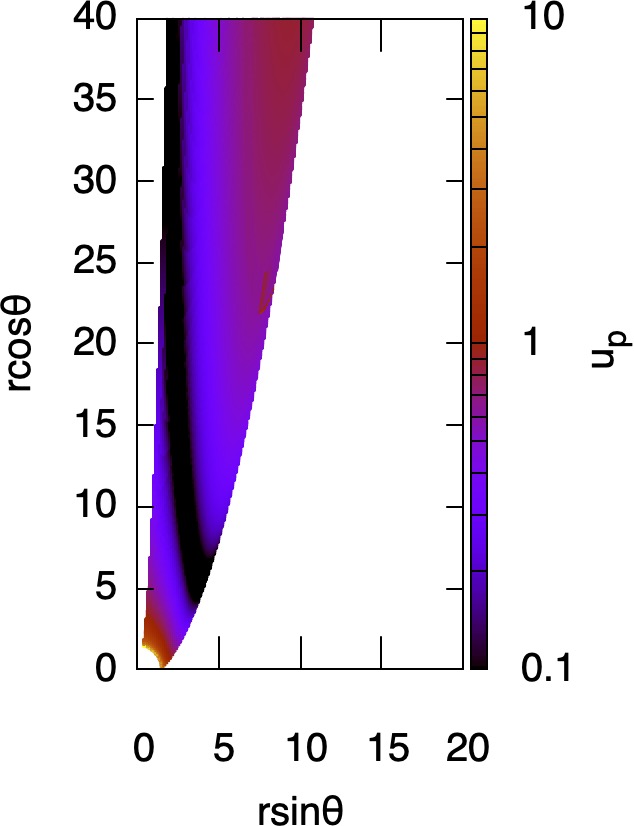}
    \includegraphics[height=0.4\textwidth]{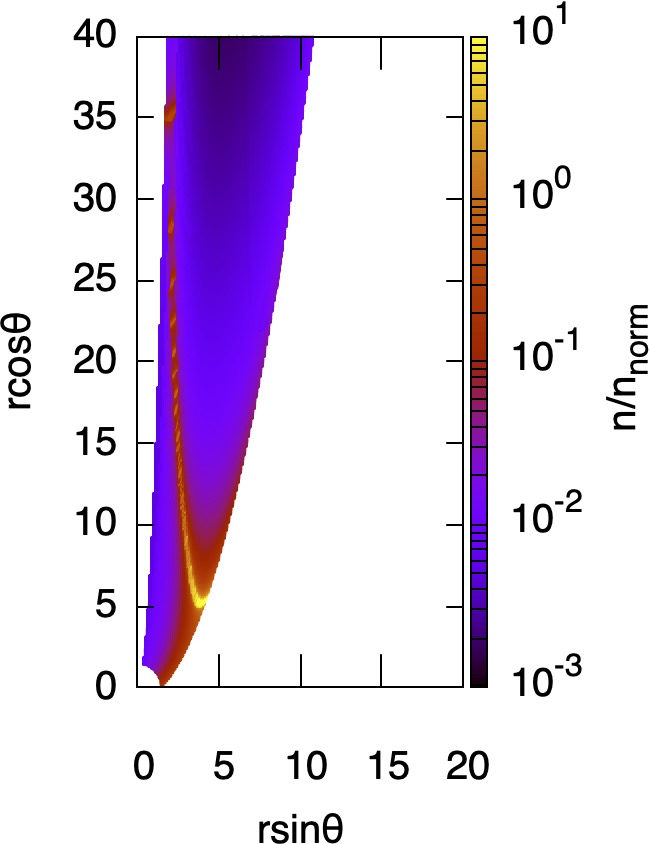}
    \includegraphics[height=0.4\textwidth]{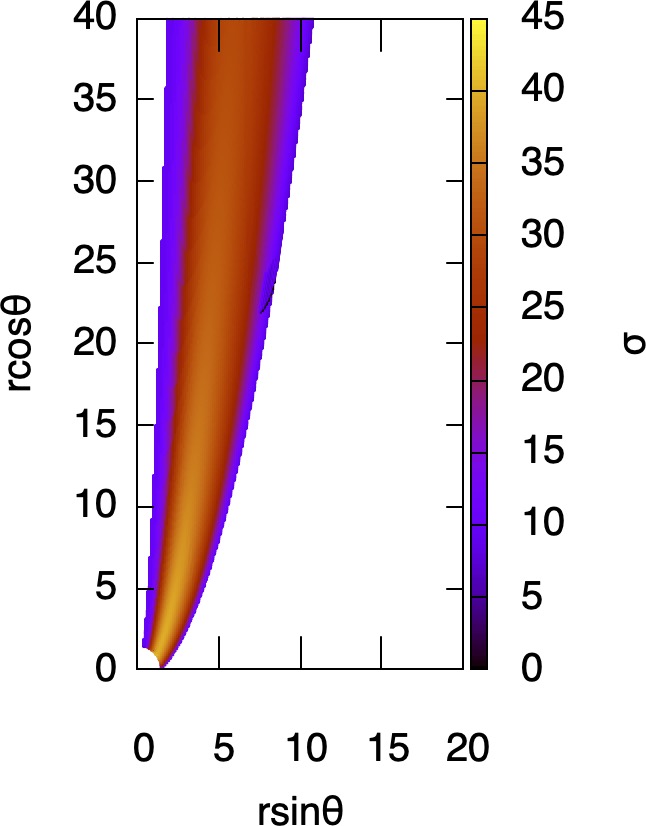}
    \caption{Two dimensional distribution of $u_p$, $n/n_{norm}$, and $\sigma$ of the P1 model.}
    \label{fig:parabolic 2d}
\end{figure*}

\begin{figure*}
    \centering
    \plottwo{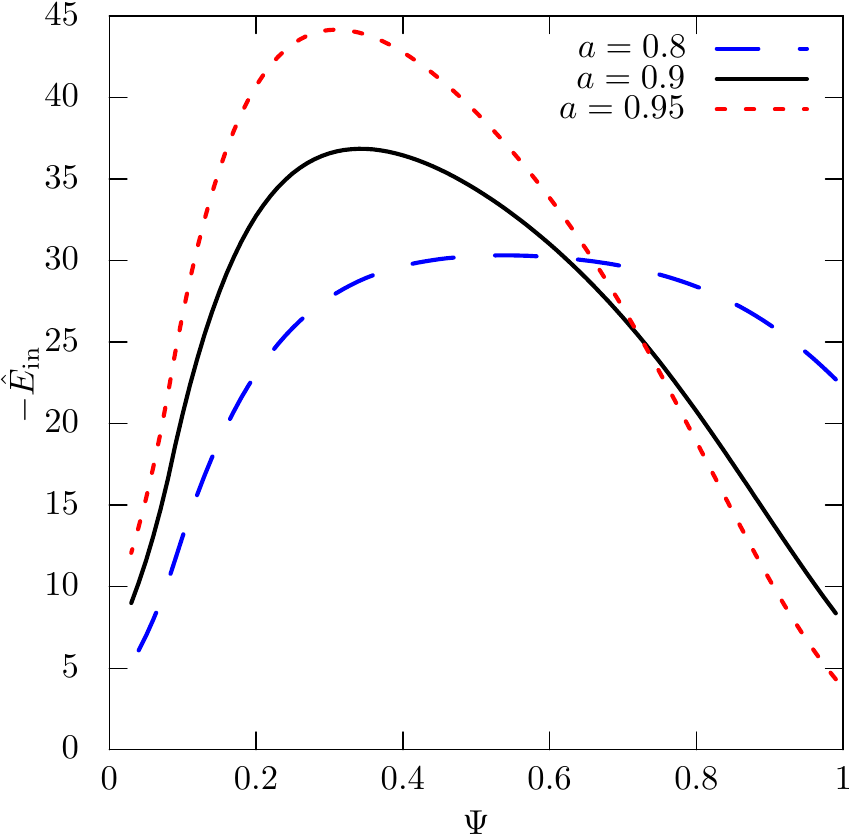}{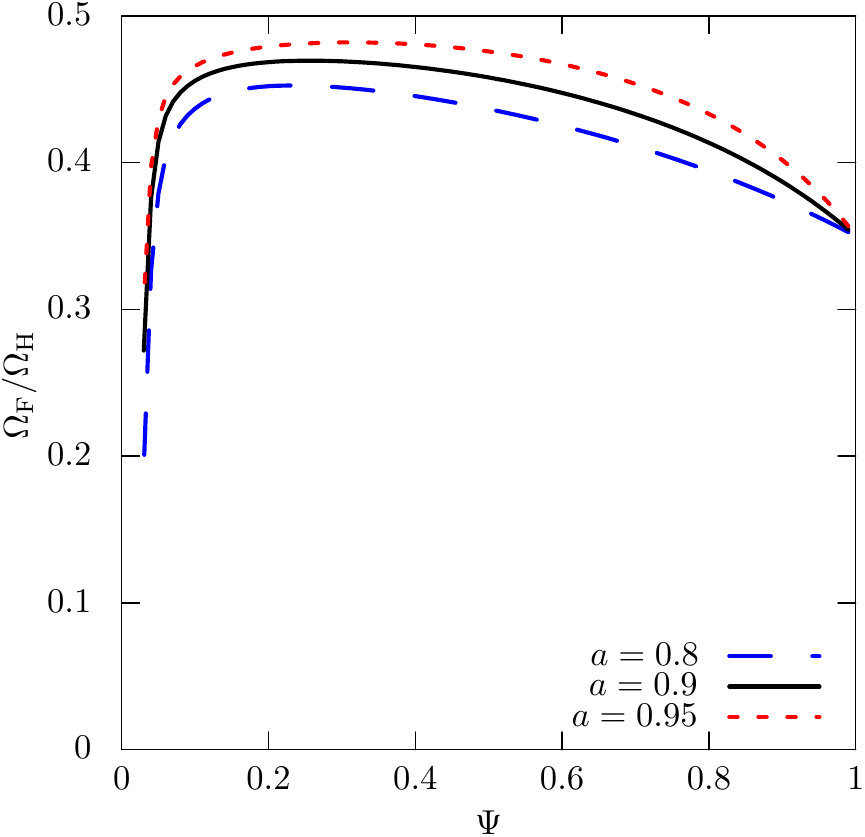}
    \plottwo{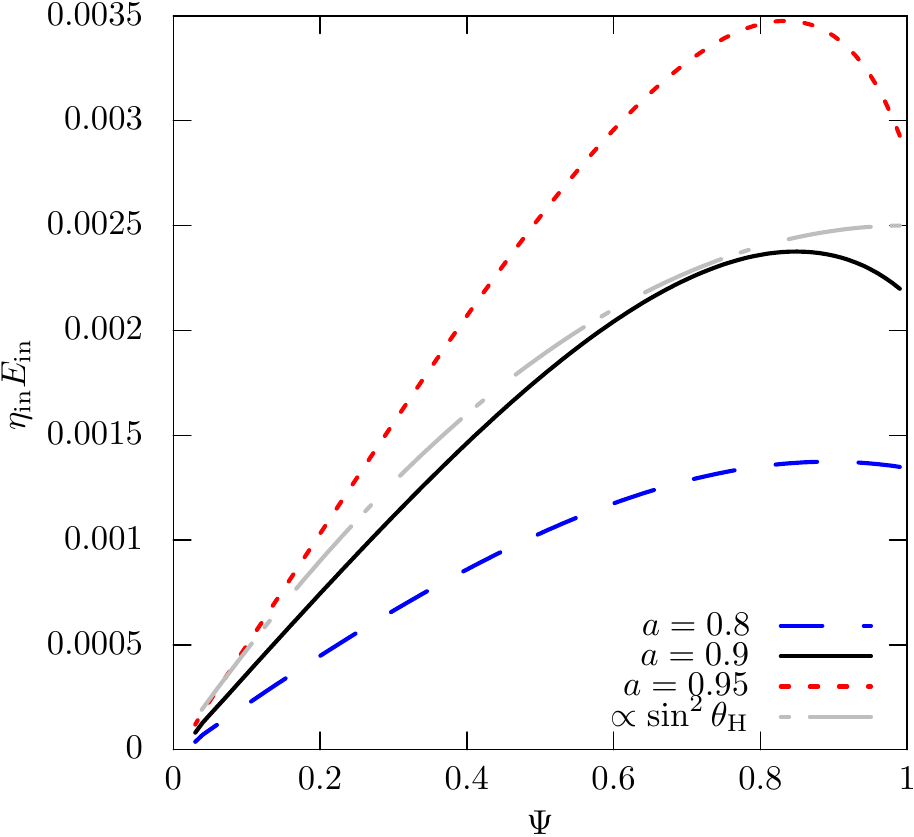}{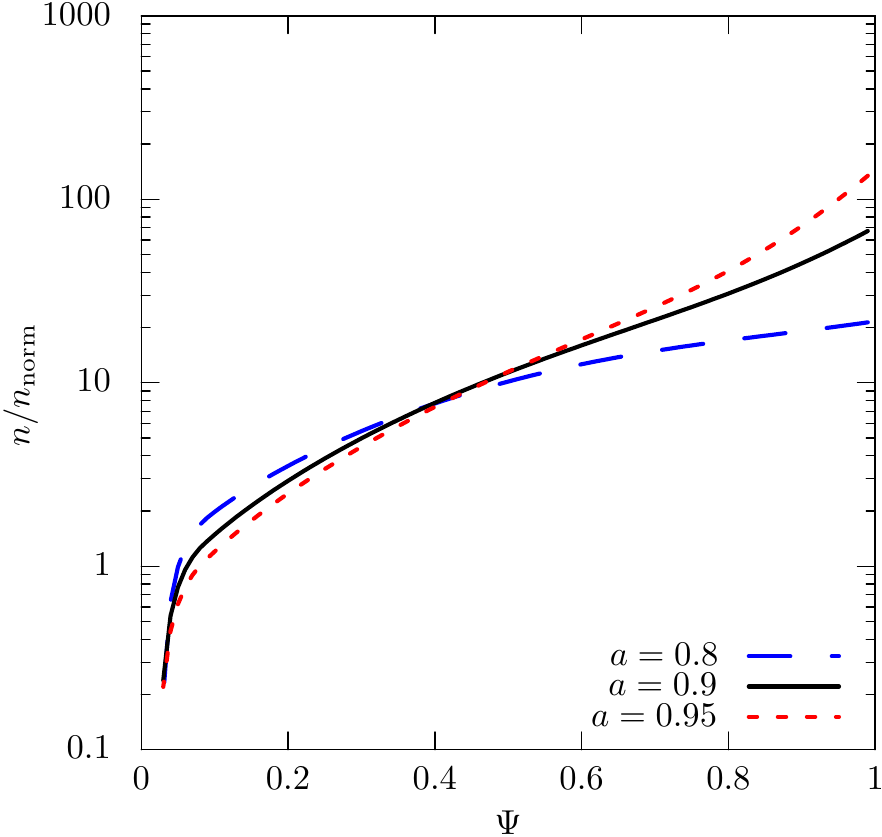}
    \caption{$\hat E_{in}$, $\Omega_F$, $\eta_{in} E_{in}$, and $n/n_{norm}$ at the separation surface. The black, blue, and red lines are the results of P1, P2, and P3 models, respectively. The number density at the separation surface is concentrated more near the jet edge when the BH spin $a$ is larger.}
    \label{fig:parabolic ss}
\end{figure*}

\subsection{Split-monopole configuration model} \label{sec:split-monopole configuration case}

Figure \ref{fig:monopole ss} shows $\Omega_F(\Psi)$ and $\hat L_{\rm out} \Omega_F / \hat E_{\rm out} (\Psi)$ at the separation surface. 
$\Omega_F/\Omega_H=0.5$ and the force-free condition $\hat L_{\rm out} \Omega_F / \hat E_{\rm out}=1$ are satisfied within 1\% accuracy in $\Psi>0.1$, which mean that our approximate solutions are consistent with the force-free monopole solution \citep{Blandford1977}.
The deviation from the monopole force-free solution decreases,
as either the BH spin is smaller, $\hat E_0$ is larger, $u_{p,ss}$ is smaller, or $\Omega_F(\Psi=1)$ is closer to $0.5\Omega_H$.  

Figure \ref{fig:monopole wind} shows $u_p$ and $B_3$ along the field lines of $\Psi=0.1, 0.5, 0.9$.
The outflow does not pass through the fast magnetosonic point, unlike in the parabolic field configuration case, as discussed in \citet{Camenzind1986a}. 
$B_3$ of each flow is almost constant along the field line unless it diverges. 
This means that the conversion from the Poynting flux to the fluid energy flux is inefficient in the monopole field configuration and that the electromagnetic field is almost force-free in the whole region.

\subsection{Parabolic configuration model} \label{sec:parabolic configuration case}

\subsubsection{P1 model}
In this subsection, we focus on the results of the calculation of the P1 model.
We show the two dimensional distribution of $u_p$, $n$, and $\sigma$ in Figure \ref{fig:parabolic 2d}. 
Here, we normalize the number density by 
\begin{equation}
    n_{norm} \equiv \left[ \frac{B_1 B^1 + B_2B^2 + B_3B^3}{8\pi\mu} \right]_{(r=r_{ss},\Psi=1)}.
\end{equation}
The inflow and outflow smoothly accelerate from the separation surface to relativistic speeds, and the density decreases with the distance from the separation surface.
We note that the density does not diverge at the separation surface since $u_{p,ss}$ is not zero.
At the separation surface, $\sigma$ has the almost same value as $\hat E(\Psi)$ because $|u_0|\approx 1$ (see the top left panel of Figure \ref{fig:parabolic ss}). Along the separation surface, it increases from $\approx 10$ at the jet edge to the maximum value $\approx 40$, and then decreases to $\approx 10$ near the axis. 
Along the field line, $\sigma$ decreases as the radius increases for both the inflow and outflow.

Figure \ref{fig:parabolic ss} shows $\eta(\Psi) E(\Psi)$, $\hat E(\Psi)$, $\Omega_F(\Psi)$, and $n/n_{norm}$ at the separation surface.
$\hat E(\Psi)$ has a peak at $\Psi \sim 0.3$.
The Poynting flux becomes zero at the axis, which means $\hat E(\Psi=0) = -u_0 \approx 1$. 
$\Omega_F$ increases toward $\Omega_F=0.5\Omega_H$ from the edge to the axis, but it decreases near the axis at $\Psi \approx 0.25$. 
$\hat E \approx \hat L \Omega_F$ is satisfied within 1\% accuracy.
This means that the flow is Poynting flux dominated at the separation surface.
$\eta E$ roughly follows $\propto \sin^2\theta_H$, while this dependence is that of $B_3$ on $\theta_H$ at the horizon for $a \ll 1$ (Equation~\ref{eq:B3}).
The number density at the separation surface has the peak at the jet edge and decreases to nearly zero toward the jet axis. 

\begin{figure}
    \centering
    \plotone{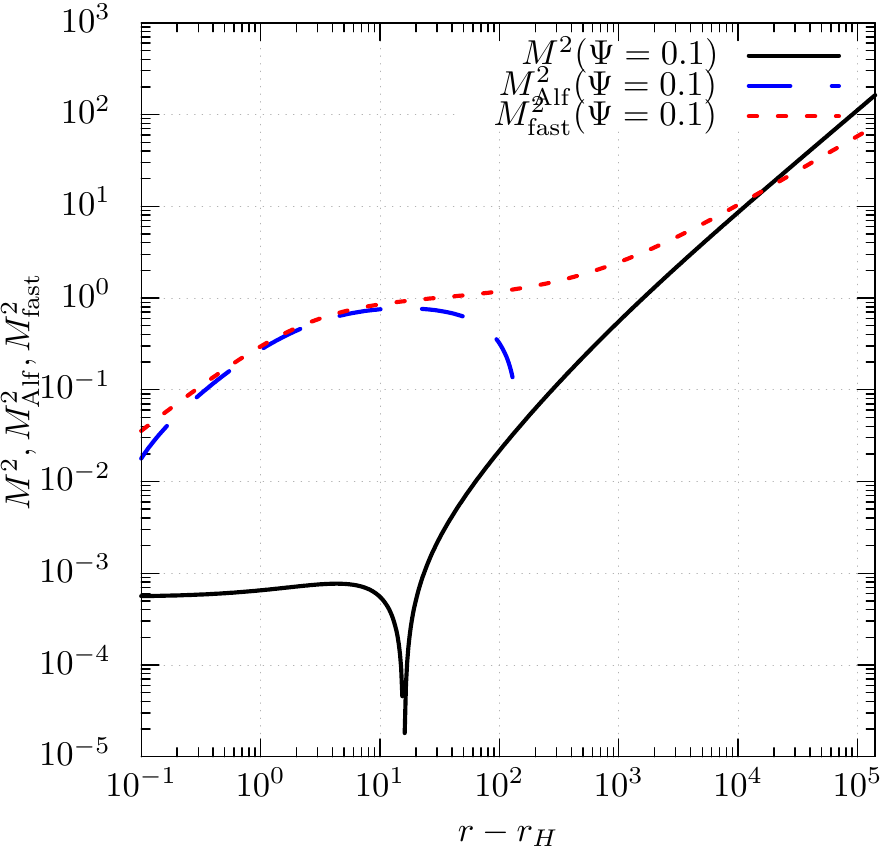}    
    \plotone{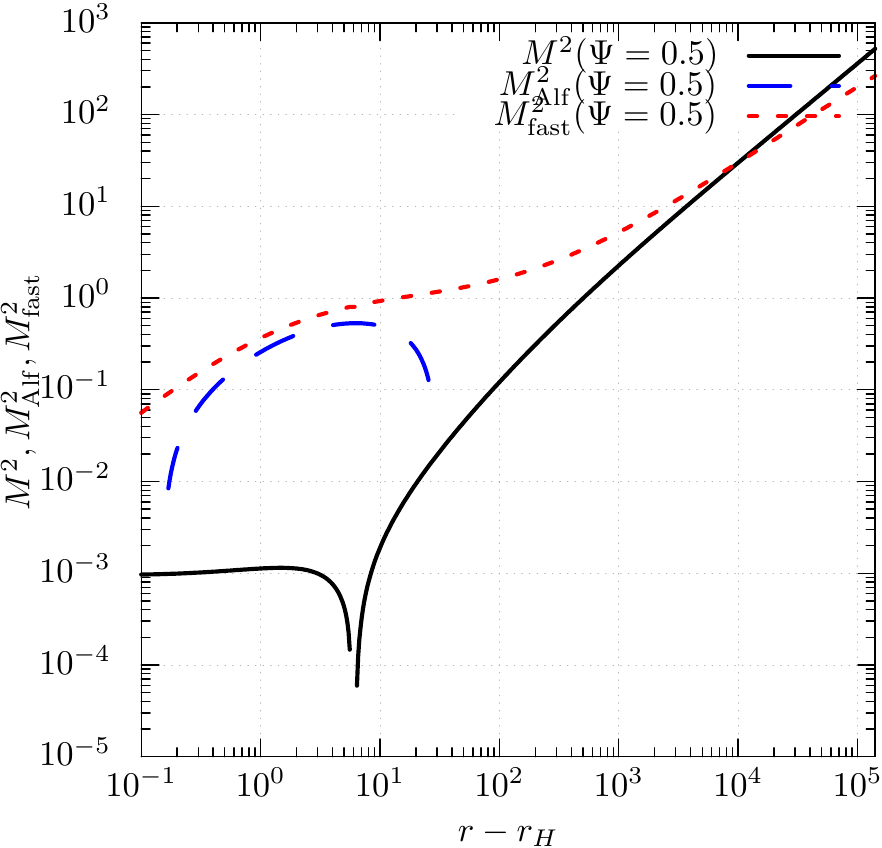}    
    \plotone{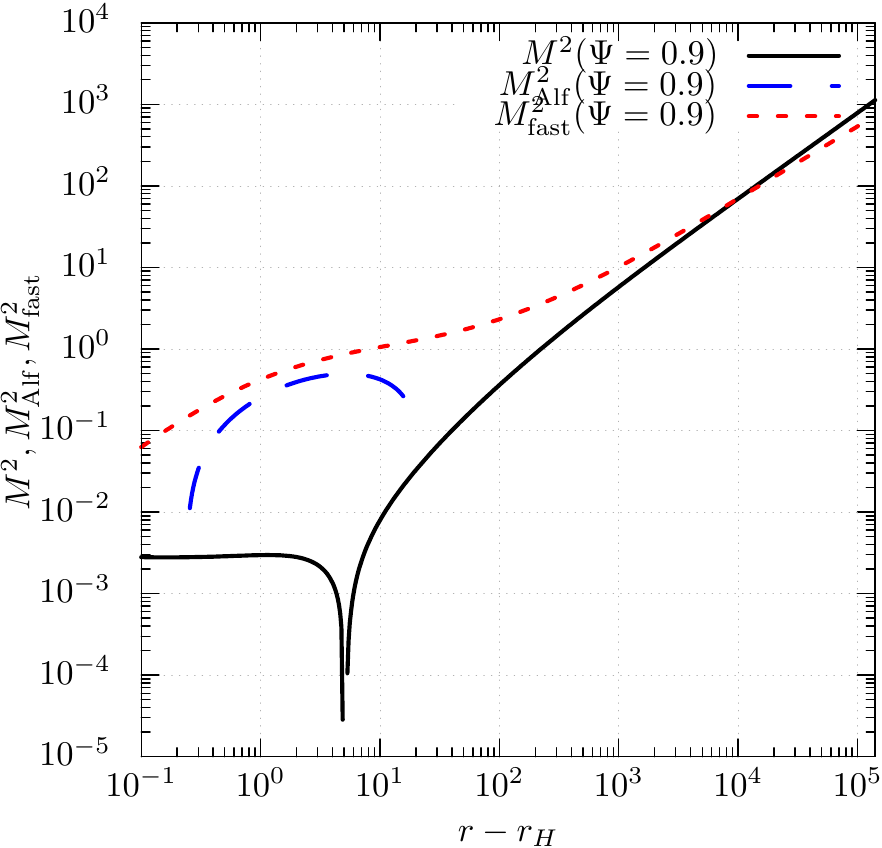}
    \caption{$M^2$, $M^2_{Alf}$, and $M^2_{fast}$ along the field lines of $\Psi=0.1, 0.5,$ and $0.9$ in the P1 model. The intersections of $M^2$ and $M^2_{Alf}$ are the Alfven points and the ones of $M^2$ and $M^2_{fast}$ are the fast magnetosonic points. }
    \label{fig:parabolic wind M2}
\end{figure}

\begin{figure}
    \centering
    \plotone{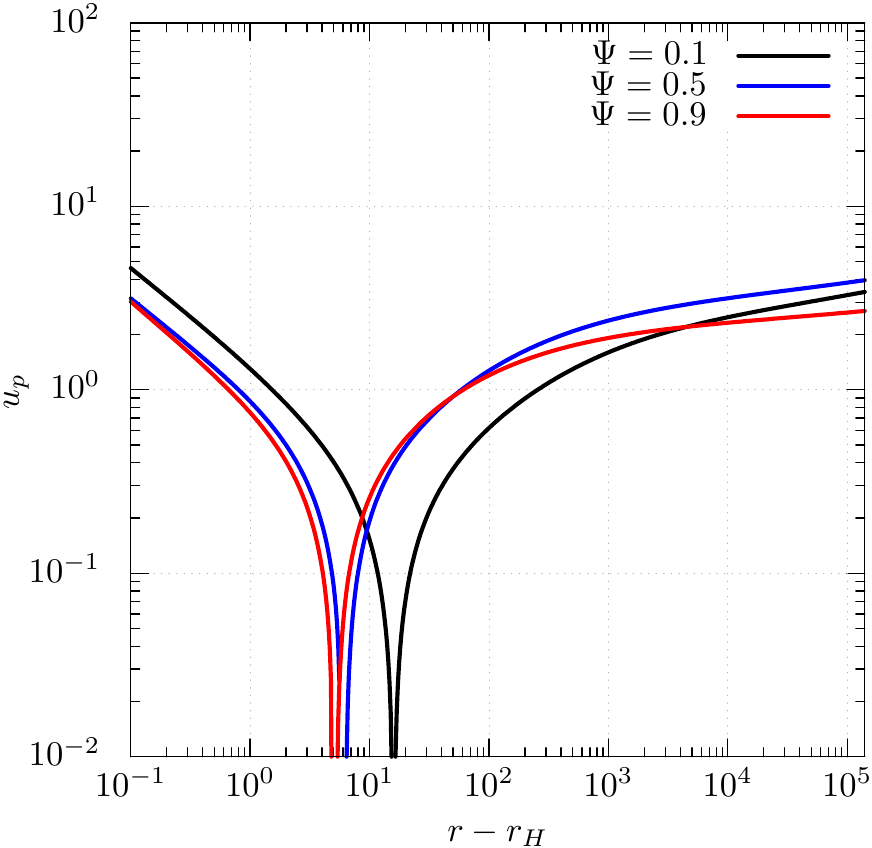}
    \caption{$u_p$ along the field lines $\Psi=0.1, 0.5,$ and $0.9$ in the P1 model. We note that $u_p$ is the absolute value of the poloidal velocity.
    }
    \label{fig:parabolic wind}
\end{figure}

Figure \ref{fig:parabolic wind M2} shows $M^2$, $M^2_{Alf}$, and $M^2_{fast}$ along the field lines of $\Psi=0.1, 0.5,$ and $0.9$, where $M^2_{Alf} \equiv k_0$.
The intersections of $M^2$ and $M^2_{Alf}$ are the Alfven points, and the ones of $M^2$ and $M^2_{fast}$ are the fast magnetosonic points.
Figure \ref{fig:parabolic wind} shows $u_p$ along the field lines $\Psi=0.1, 0.5,$ and $0.9$.
Both the inflow and outflow start from the separation surface and accelerate to relativistic speeds. 

We evaluate the trans-field force-balance by introducing 
$\chi \equiv \left| (f_+ - |f_-|) / (f_+ + |f_-|) \right|$,
where we gather the positive components of Equation (\ref{eq:trans-field EoM}) to $f_+$ and the negative ones to $f_-$.
$\chi$ ranges from 0 to 1, and $\chi = 0$ means  
the complete force balance.
The results show that $\chi$ is less than $10^{-6}$ for all the field lines at the separation surface.
For the outflow above the separation surface, $\chi$ increases rapidly to $\sim 10^{-1}$ and then turns to decrease, while for the inflow, it increases up to $\sim 1$ and then decreases.
The Lorentz force directs toward the axis $e_{(n)A} F^{A\nu}I_\nu > 0$ and the inertial force directs in the opposite way $- e_{(n)A} \rho u^A u^\nu_{;\nu} < 0$. 
These forces are well balanced at the separation surface $e_{(n)A} \rho u^A u^\nu_{;\nu} \approx e_{(n)A} F^{A\nu}I_\nu$.
In other radii than the separation surface, the Lorentz force is larger than the inertial one, and the net force is directing for the flow to collimate.

\subsubsection{Parameter dependences}
We calculate the parabolic configuration models with different parameter values listed in Table \ref{tab:parameter list} to investigate the dependencies of the density distribution on the BH spin $a$ and $u_{p,ss}$.

We use $a=0.8$ and $0.95$ for the P2 and P3 models, respectively, to investigate the BH spin dependence.
$\hat E, \Omega_F, \eta E, n/n_{norm}$ are shown in Figure \ref{fig:parabolic ss}.
Interestingly, as $a$ becomes larger, the density gets larger near the jet edge and smaller near the axis, while $\hat E$ changes in the opposite way.
$\eta E$ of the P2, P3 models also roughly follow $\propto \sin^2\theta_H$.
$\eta E$ becomes larger in all the field lines with $a$ because of the increase of $B_3(r=r_H)$ and the Poynting flux.
$\Omega_F$ also increases with $a$.
The force-freeness $1- L \Omega_F / E$ is smaller than 0.2 for all three models.
The minimum value $1- L \Omega_F / E \approx 0.01$ realizes where $\hat E$ is maximum.

\begin{figure*}
    \centering
    \plottwo{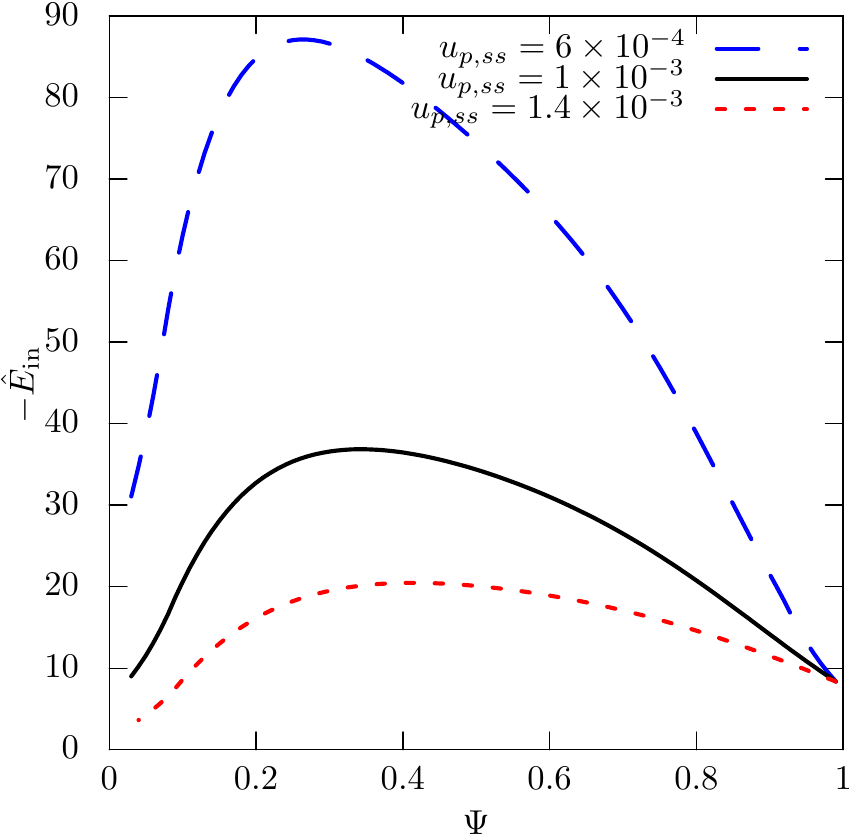}{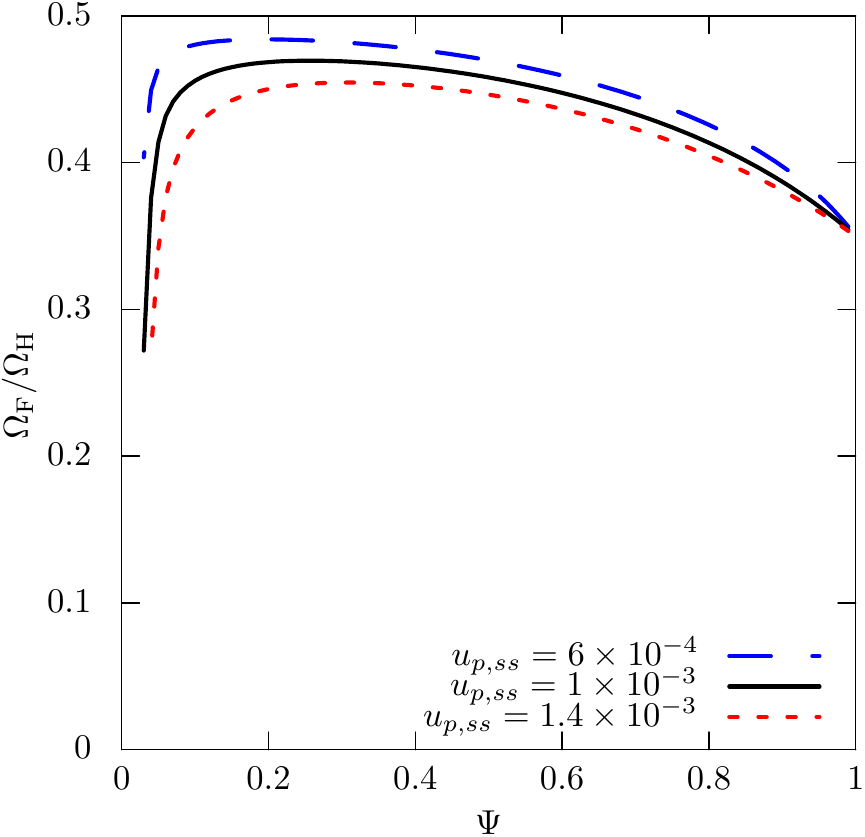}
    \plottwo{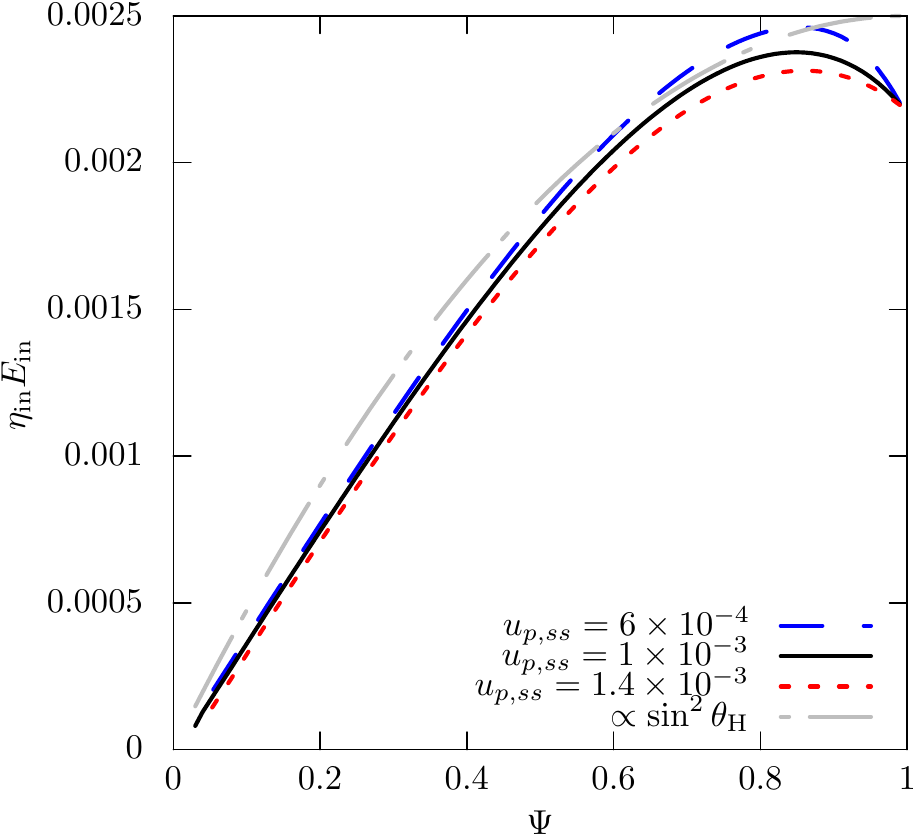}{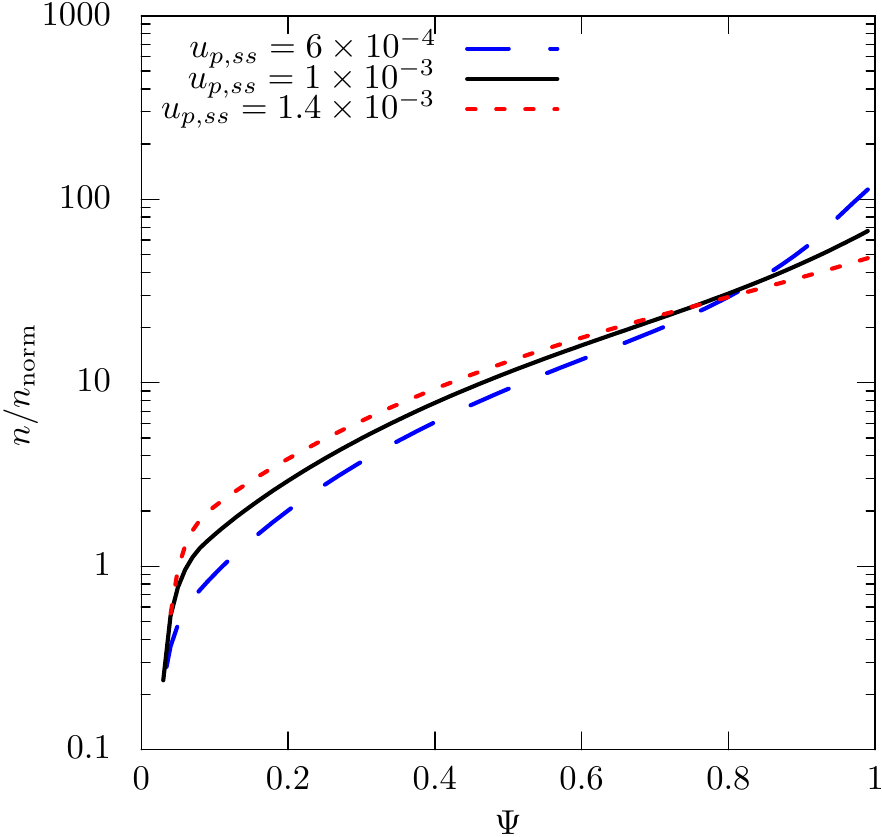}
    \caption{$\hat E_{in}$, $\Omega_F$, $\eta_{in} E_{in}$, and $n/n_{norm}$ at the separation surface. The black, blue, and red lines are the results of P1, P4, and P5 models, respectively. The number density at the separation surface is concentrated more near the jet edge when $u_{p,ss}$ is smaller.}
    \label{fig:upss dependence}
\end{figure*}

We perform calculations with different $u_{p,ss}$.
We use $u_{p,ss} = 6 \times 10^{-4}$ and $1.4 \times 10^{-3}$ for the P4 and P5 models, respectively.
The results are shown in Figure \ref{fig:upss dependence}.
When $u_{p,ss}$ is smaller, $n/n_{norm}$ changes in a similar fashion as $a$ gets larger.
$n/n_{norm}$ at the jet edge changes proportional to $u_{p,ss}^{-1}$.
The P1, P4, and P5 models show that $\eta E$ does not significantly depend on $u_{p,ss}$.
As $u_{p,ss}$ decreases, $\hat E$ increases and $\eta$ decreases for all the field lines. 

\begin{figure}
    \centering
    \plotone{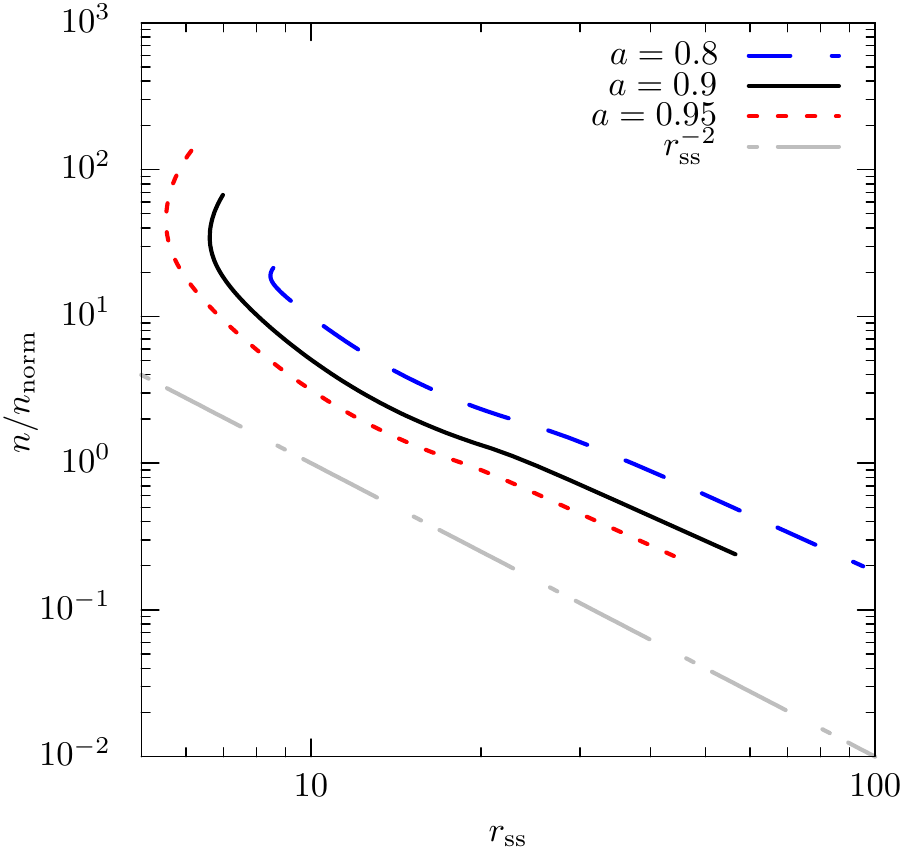}
    \caption{$n/n_{norm}$ as a function of $r_{ss}$. The black, blue, and red lines are the results of the P1, P2, and P3 models, respectively. When the BH spin is larger, $r_{ss}$ decreases. At the far zone, $n/n_{norm}$ roughly follows $\propto r_{ss}^{-2}$.}
    \label{fig:density distribution along the separation surface}
\end{figure}

\begin{figure}
    \centering
    \plotone{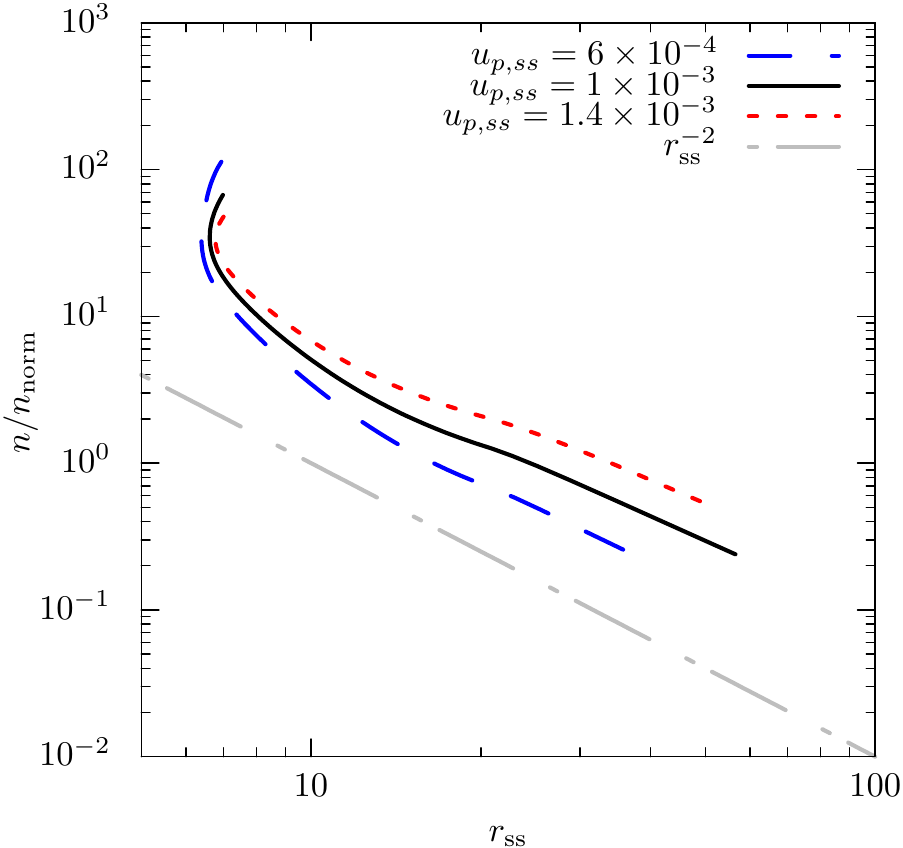}
    \caption{$n/n_{norm}$ as a function of $r_{ss}$. The black, blue, and red lines are the results of the P1, P4, and P5 models, respectively. When the BH spin is larger, $r_{ss}$ decreases. At the far zone, $n/n_{norm}$ roughly follows $\propto r_{ss}^{-2}$. }
    \label{fig:rss-n upss dependence}
\end{figure}

\section{discussion} \label{sec:discussion}

\subsection{Density on the separation surface  }
The matter density distribution will constrain the mass-loading mechanism.
In Figure \ref{fig:density distribution along the separation surface}, we show $n/n_{norm}$ of the P1, P2, and P3 models as a function of $r_{ss}$.
$n/n_{norm}$ is largest at the jet edge and decreases as $\Psi$ get smaller as shown in Figure \ref{fig:parabolic ss}.
$r_{ss}$ decreases as $a$ gets larger.
At the far zone, the normalized density roughly follows $n/n_{norm} \propto r_{ss}^{-2}$ in all the models.
Figure \ref{fig:rss-n upss dependence} shows $n/n_{norm}$ as a function of $r_{ss}$ for the different values of $u_{p,ss}$. 
We also have the dependence $n/n_{norm} \propto r_{ss}^{-2}$ in the far zone in these models.
For $r_{ss}<10$, the $r_{ss}$ dependence of $n/n_{norm}$ is steeper.

The annihilation of high-energy photons from the accretion disk is one of the proposed mass-loading mechanisms \citep{Levinson2011, Moscibrodzka2011, Kimura2020}. 
This process leads to the $e^+ e^-$ density distribution $n  \propto r^{-6}$ for the case in which $\gamma$-ray emitting region is compact near the BH \citep{Moscibrodzka2011}, and then the particles are injected mainly at the base for the outflow, i.e., at the separation surface, which is similar to the situation of our model. However, the dependence $r^{-6}$ appears too steep compared with the results in our model.
If $\gamma$-ray emitting region is extended, say around $r \sim 10$, in the accretion flow \citep{Kimura2020}, the $r$ dependence of the $e^+e^-$ density can be much shallower and might be consistent with our results.
However, it should be noted that this $e^+e^-$ injection model can provide particle number density sufficient for screening the spark gap (i.e., larger than the Goldreich-Julian number density), but not sufficient for the radio synchrotron flux of M87 jet \citep{Kimura2020}. 
Other injection mechanisms such as magnetic reconnection \citep[e.g.][]{Parfrey2015, Mahlmann2020} and/or fluid instability \citep[e.g.][]{Globus2016, Nakamura2018, Chatterjee2019, Sironi2020} could be efficient for injecting electrons that produce the limb-brightened radio emission.

The results of our model are compatible with the observed limb-brightened emission structure of jets (see also Figure~\ref{fig:parabolic 2d}), although it is uncertain what fraction of the matter contribute to the non-thermal emission.  
The relative amount of the density near the axis compared to the one near the jet edge is also important because the emission near the axis will be Doppler-boosted, forming the central ridge of the observed triple-ridge emission structure of M87 jet \citep{Ogihara2019}.

\subsection{Comparison to other studies}
The distribution of the Bernoulli parameters can be compared to the results of other studies.
\citet{Huang2020} numerically solved the GS equation for the whole region. They set the loading zone between the separation surface and the null-charge surface, although it is unclear whether such a large loading zone is necessary for obtaining the solutions.
$\hat E(\Psi)$ has a peak near the jet edge in their result, while our models have the one relatively closer to the axis.
They showed that $\Omega_F$ monotonically increases toward the axis, and $\Omega_F(\Psi=0)=0.5\Omega_H$, which is set as a boundary condition, while in our model, $\Omega_F(\Psi)$ decreases rapidly near the axis. 
$\eta$ of their model is assumed by a given magnetization parameter and the poloidal magnetic field at the null-charge surface. It is larger near the axis, which is the opposite trend from our results.

$\Omega_F(\Psi)$ distribution is also shown in \citet{Beskin2013}, in which they solve the GS equation of a cylindrical jet in a special relativistic regime.
In their result, $\Omega_F(\Psi)$ decreases near the axis like our results. 
This trend is also seen in the GRMHD simulation of \citet{McKinney2012}.

\subsection{Magnetic bending profile}
\citet{Pu2020} introduced the reasonable shape of the function of $E_{p,ZAMO}/B_{T,ZAMO}$, which can be rewritten by the bending angle of the field line, and derived wind solutions with the prescribed function \citep[see also][]{Tomimatsu2003, Takahashi2008}.  
$E_{p,ZAMO} = |G_\phi B_p/(G_t \sqrt{g_{33}})|$ is the poloidal electric field strength and $B_{T,ZAMO} = |B_3/(\alpha \sqrt{g_{33}})|$ is the toroidal magnetic field strength in the zero angular momentum observer frame. 

There are some constraints to this function. 
First, $E_{p,ZAMO}/B_{T,ZAMO}=1$ at the horizon (i.e. the Znajek condition). 
Second, $E_{p,ZAMO}/B_{T,ZAMO}=0$ at the null-charge surface, where the field line corotate with the spacetime ($-g_{03}/g_{33}=\Omega_F$).
Finally, $(E_{p,ZAMO}/B_{T,ZAMO})^2<1-1/\hat E^2$ for the outflow in order to prevent $u_p$ from diverging. 
For the outflow, $E_{p,ZAMO}/B_{T,ZAMO}$ needs to increase for the flow to accelerate.
Additionally, $E_{p,ZAMO}/B_{T,ZAMO}$ should be smooth and continuous.

\begin{figure}
    \centering
    \plotone{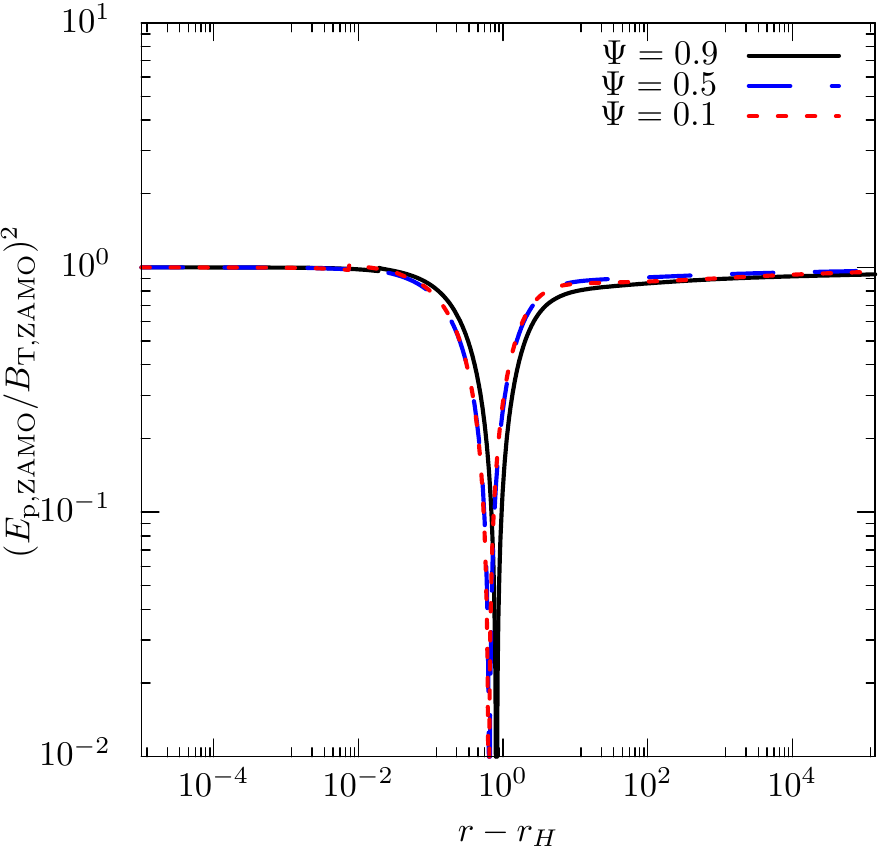}
    \caption{The radial profiles of $(E_{p,ZAMO}/B_{T,ZAMO})^2$ of $\Psi=0.1, 0.5,$ and $0.9$ in the P1 model. They are smooth and continuous and satisfy $(E_{p,ZAMO}/B_{T,ZAMO})^2 = 1$ at the horizon from the Znajek condition, $(E_{p,ZAMO}/B_{T,ZAMO})^2 = 0$ at the null-charge surface, and $(E_{p,ZAMO}/B_{T,ZAMO})^2<1-1/\hat E^2$ in the outflow region. }
    \label{fig:EPBTZAMO}
\end{figure}

The previous studies mentioned above prescribed $(E_{p,ZAMO}/B_{T,ZAMO})^2$ as a constant value for outflow and derived $u_p$ using it.
The assumed $(E_{p,ZAMO}/B_{T,ZAMO})^2$ and the derived one using Equation (\ref{eq:BB3}) was not self-consistent.
We solved Equation (\ref{eq:fourth-order up}) which does not explicitly depend on $B_3$, and derive $(E_{p,ZAMO}/B_{T,ZAMO})^2$ afterwards.
We show the self-consistent profile of $E_{p,ZAMO}/B_{T,ZAMO}$ of $\Psi=0.1, 0.5$,and $0.9$ in Figure \ref{fig:EPBTZAMO}. 
$(E_{p,ZAMO}/B_{T,ZAMO})^2$ in our result follows all the conditions listed above. 

\subsection{Model limitations}
The inflows do not pass through the fast magnetosonic point in our results. 
The inflow diverges at the region very close to the horizon.
Additionally, the indicator of the force-balance between the field lines $\chi$ is also large in the inflow region.
We try to find different values of $\hat{E}_{in}$ from those satisfying the trans-field force-balance for the inflows to pass through the fast magnetosonic points, and find that $|\hat{E}_{in}|$ needs to be smaller only by a few in the case of the P1 model. 
The difference of $E_{in}$ is smaller for larger $\Psi$.
Adjustment of $\Psi(r,\theta)$ should be considered in future work.

We note that if more particles are injected at regions closer to the BH, like in the case of the annihilation of high-energy photons, the inflows are highly affected by mass loading and should be modeled with varying $\eta$ along a field line \citep[cf.][]{Globus2014}, while the current model of outflows is still applicable.

The distribution of $u_{p,ss}(\Psi)$ may differ between the injection mechanisms. 
In our model, we assume that $u_{p,ss}(\Psi)$ is constant for all the field line. 
If we change the distribution, it may change the density profile along the separation surface.
This will be discussed in a separate work.
Injection via instabilities and the magnetic reconnection may occur by the interaction between the jet edge and the disk wind \citep[][]{Globus2016, Nakamura2018, Chatterjee2019, Sironi2020}. 
This may change the field line configuration and the density distribution near the jet edge. 

Although our current approximate solutions of GRMHD jets have the limitations mentioned above, they are useful not only for constraining the mass-loading mechanism but also for obtaining key ingredients from the observed complex emission structures. Indeed, by comparing the approximate special relativistic force-free steady jet models with observations of far-zone jets, it is found that the special relativistic beaming effect is essential for understanding the symmetry of the observed limb-brightened structure and the central ridge emission and that the emission image strongly depends on the matter density distribution at the base of outflow \citep{Takahashi2018, Ogihara2019}. This has motivated us to develop GRMHD steady jet models.

\section{summary \& prospects} \label{sec:summary}
We constructed a steady, axisymmetric GRMHD jet model to examine the density distribution inside the jet. 
We fixed the poloidal field line configuration, which mimics the ones of force-free or GRMHD simulation results with an additional term for obtaining trans-fast-magnetosonic outflows,
and assumed the poloidal velocity of the flow at the separation surface $u_{p,ss}$ as constant for different field lines.
We numerically solved the trans-field force-balance (Equation \ref{eq:trans-field EoM}) at the separation surface to determine the Bernoulli parameters and analytically solved the fourth-order equation of the poloidal velocity along the field lines (Equation \ref{eq:fourth-order up}).

We consider the two field line configurations; (i) the split-monopole configuration model and (ii) the parabolic configuration model. 
(i) In the split-monopole configuration model, we confirmed that our method could reproduce the BZ solution well except around the axis. 
(ii) In the parabolic configuration model, we obtained approximate solutions of highly-magnetized GRMHD flows.
The force-balance between field lines at the separation surface is satisfied with high accuracy. The outflow successfully passes the fast magnetosonic point and satisfies a good force-balance.
We found that the number density distribution at the separation surface 
roughly scales as $n(r=r_{ss}, \Psi) \propto r_{ss}^{-2}$ in the far zone.
We examined the parabolic configuration models with the different BH spins $a$ and $u_{p,ss}$.
As the BH spin increases, the density at the separation surface is concentrated more at the jet edge. 
We obtained the similar trend when $u_{p,ss}$ is smaller.

Our semi-analytic model can be utilized to examine the density distribution of the highly magnetized jet region, while current GRMHD simulations cannot since they use the artificial density floor, and can survey a larger parameter space in detail because of the smaller computational budget.
It will be interesting to apply radiative transfer calculations to our model and compare them to the observed emission structure, as done with the special relativistic force-free jet models \citep{Takahashi2018, Ogihara2019}.

Observations of core shifts, collimation profiles, proper motions of blobs, and Faraday rotation maps in various frequencies and scales \citep[][]{Kovalev2007, Asada2013, Hada2011, Hada2013, Hada2018, Nakamura2013, Mertens2016, Park2019, Park2019a, Kim2020, Akiyama2018, Hiura2018, Giovannini2018, Kravchenko2020} will also be useful for testing jet models \citep{Kino2014, Kino2015, Chael2019, Jeter2020}.
Future EHT and EATING-VLBI observations will provide crucial information on the jet launching region of M87 \citep{Hada2020}.

\acknowledgments
We thank the anonymous referee for useful comments. We thank Shigeo Kimura, Hung-Yi Pu, Fumio Takahara, and Masaaki Takahashi for fruitful discussions.
T.O. acknowledges financial support from the Graduate Program on Physics for the Universe of Tohoku University. 
This work is partly supported by Grant-in-Aid for JSPS Fellowship 20J14537 (T.O.) and KAKENHI No. 18H01245 (K.T.). 

\end{CJK*}

\bibliography{ref}

\end{document}